\documentstyle[12pt]{article}

\newcommand{\gwig}{\mbox{\,\raisebox{.3ex}
{$>$}$\!\!\!\!\!$\raisebox{-.9ex}{$\sim$}}\,}
\begin{document}
\baselineskip=18pt

\begin{titlepage}
\begin{flushright} UCLA/TEP/93/35 \\
October 1993
\end{flushright}
\vspace{0.75cm}
\begin{center}
{\bf \LARGE Higgs Boson Interference in
$\gamma \gamma \rightarrow W^+W^-$}
\vspace{0.75cm}

{\bf D.A. Morris,$^a$\footnote{email: morris@uclahep.bitnet}
T.N. Truong$^b$\footnote{email: pthtnt@frpoly11}
and D. Zappal\'{a}$^a$\footnote{email: zappala@uclahep.bitnet}}
\\~~ \\
$^a$University of California, Los Angeles\\
405 Hilgard Ave., Los Angeles, CA, 90024 U.S.A. \\ ~~ \\
$^b$Centre de Physique Th\'{e}orique de l'Ecole Polytechnique,\\
91128 Palaiseau, France
\date{}
\end{center}
\vspace{.5cm}
\thispagestyle{empty}
\begin{abstract}\normalsize
\noindent
We study interference effects between resonant
and nonresonant amplitudes for the
$\gamma \gamma \rightarrow W^+ W^-$ process at
a backscattered photon-photon collider.
We show that a Higgs boson with $M_H \gwig 200~{\rm GeV}$
is manifest as a resonant dip in the $W^+W^-$ invariant mass
spectrum and we investigate its statistical significance.
\end{abstract}
\vspace{2.cm}

\vspace{1ex}
\end{titlepage}

There has been recent interest in weak gauge boson pair production
in $\gamma \gamma$ collisions as a means of testing the Standard Model
and studying a Higgs boson with a mass up to a few
hundred GeV\cite{gu92,bow93,bo93}. Photons with the necessary energies
may be produced by backscattering low energy laser beams
at future $e^+e^-$ or $e^-e^-$ facilities\cite{bo93,gi83,tel90}.
In this letter we examine the prospects for
studying a Higgs boson with $M_H > 2 M_Z$ through the process
$\gamma \gamma \rightarrow H \rightarrow W^+W^-$
by investigating the role of quantum interference effects.
\pagestyle{plain}

To lowest order, $\gamma \gamma \rightarrow H \rightarrow W^+W^-$
proceeds through the one-loop diagrams of Fig.~1a.
However, due to a large background of nonresonant tree-level processes
(see Fig.~1b), it has generally been concluded that the $W^+W^-$ decay
mode of a heavy Higgs boson
would be difficult to exploit. $H \rightarrow Z Z,$
which has no tree-level background, looks more promising in this
respect\cite{gu92,bo93,ji93,be93,di93,ve93}.
Unfortunately, cursory studies of the
$H \rightarrow W^+W^-$ channel have neglected interference between
the tree and loop amplitudes of Fig.~1: constructive interference
effects could strengthen the viability of the $W^+W^-$ mode.
In this letter we report our findings of relatively
large destructive interference effects in the $\gamma \gamma \rightarrow
W^+W^-$  channel which imply that, in marked contrast to earlier studies,
a Higgs boson of mass $M_H \gwig 200~{\rm GeV}$
is manifest as a resonant dip in the $W^+W^-$ invariant mass
spectrum.

We are interested in the process
$\gamma \gamma \rightarrow W^+ W^-$ in vicinity of
a Higgs boson resonance
 (i.e., $ | M_{WW} - M_H | / \Gamma_H
\simeq {\cal O}(1)$ where $M_{WW}$ is the $W^+W^-$ pair
invariant mass and $\Gamma_H$ is the Higgs boson decay width).
The relevant diagrams are shown in Fig.~1. For $M_{WW} = M_H$
the loop amplitudes of Fig.~1a are of $\cal{O}(\alpha)$
due to a cancellation of weak coupling constants between the
$HW^+W^-$ vertex, the $Hf\bar{f}$ vertex and $\Gamma_H$
in the Higgs boson propagator.
For $| M_{WW} - M_H | / \Gamma_H \gg {\cal O}(1)$
the loop amplitudes are ${\cal O}(\alpha \alpha_w )$
so that, strictly speaking, a complete calculation in this region
should include ${\cal O}(\alpha \alpha_w )$ contributions from one-loop
radiative corrections to the nonresonant tree amplitudes of Fig.~1b.
We can safely neglect such corrections for our purposes
since they are ${\cal O}(\alpha_W)$ smaller than the tree-level amplitudes
and they vary smoothly over the interval $ | M_{WW} - M_H | / \Gamma_H
\simeq {\cal O}(1)$ where interference phenomenon occur.

        The literature contains separate calculations for the helicity
amplitudes corresponding to the loop diagrams of Fig. 1a \cite{va79}
and the tree diagrams of Fig.~1b \cite{ye91,be92}. Unfortunately,
since such calculations do not chronicle their choice of phase
conventions, it is impossible to combine the results of different authors
with any degree of certainty. We have recalculated
the helicity amplitudes of Fig.~1 using uniform phase conventions
and find complete agreement (modulo different phase transformations)
with the amplitudes of Refs.~\cite{va79,be92}\footnote{Due
to an apparent inconsistency in the labeling of polarization vectors for
transverse $W$ bosons in Ref.~\cite{ye91}, our tree
amplitudes differ from those of Ref.~\cite{ye91} by more than
a simple phase convention.}. For completeness we
list the relevant phase conventions and helicity amplitudes in the Appendix.

Fig. 2 shows the $\gamma\gamma \to W^+W^-$ cross
section $\sigma_{\lambda\lambda'}$
for various  Higgs boson masses as function of the $\gamma \gamma$
center of momentum energy
$\sqrt{s_{\gamma \gamma}}$ where $\lambda,\lambda' = \pm $
denote the photon helicities. The cross sections are integrated
over all scattering angles and summed
over the polarization states of the $W$ bosons.
Though the overall magnitude of resonance phenomena
decreases as $M_H$ increases (due to $\Gamma_H$ increasing), there
is a trend suggesting overall destructive interference
as $M_H$ increases. Throughout this letter we use $M_Z= 91.17~{\rm GeV},
M_W = 80.22~{\rm GeV}, m_t = 150~{\rm GeV}$ and neglect threshold
effects in the Higgs boson width and in
the direct production of W boson pairs.

Let us investigate the extent to which
the interference features of Fig.~2 survive
in the environment of a $\gamma \gamma$ collider.
Assuming that each $e^+(e^-)$ gives precisely one backscattered
photon, the $W^+W^-$ invariant mass spectrum at a backscattered
$\gamma \gamma$ facility is given by
\begin{equation}
\label{eq:weighted}
\displaystyle{ d\sigma \over dM_{WW} }   \,\, =
{\displaystyle 2 M_{WW} \over s_{e^+ e^-} }
{\displaystyle \sum_{\lambda \lambda'} } \,\,
 {\displaystyle \int }
{\displaystyle dx \over x }\, \, d\!\cos\theta^*\,\,
f_{\lambda}(x) \,\,
f_{\lambda'}\!\!\left(
{\displaystyle M_{WW}^2 \over  x s_{e^+e^-} } \right)  \,
{\displaystyle d\sigma_{\lambda \lambda'}\over
d\!\cos\theta^*}  \,\, ,
\end{equation}
where  $f_\lambda(x)$ is the probability
that laser photon backscatters to become a
photon with helicity $\lambda$ carrying a
fraction $x$ of the beam energy $\sqrt{s_{e^+e^-}}/2.$
The precise form of $f_\lambda(x)$ depends on the degree of
polarization of both the laser and the $e^+(e^-)$ beams.
Efficiency considerations suggest\cite{tel90} that the laser energy
$\omega$ and the  $e^+(e^-)$ beam energy be chosen in a manner such that the
limits of integration become
\begin{equation}
x_{max} =
{ \displaystyle 2 \omega \, \sqrt{s_{e^+e^-} } \over 2
\omega \sqrt{s_{e^+e^-}} + m_e^2
  } = .828 ,\qquad \qquad x_{min} =
{\displaystyle M_{WW}^2 \over x_{max} \, s_{e^+e^-}  }\, \, \, .
\end{equation}
The limits of integration over the $W^+W^-$ center of mass
scattering angle $\theta^*$ are functions of $x$ and
laboratory acceptance criteria.

For purposes of illustration, let us restrict our attention to an
experimental arrangement in which completely circularly polarized
laser photons (with helicities chosen so that \addtocounter{footnote}{2}
$\lambda^{\rm laser}_1 = \lambda^{\rm laser}_2$)\footnote{The case
$\lambda^{\rm laser}_1 \ne \lambda^{\rm laser}_2$ gives less promising
results which we do not present.}
backscatter off completely unpolarized $e^+e^-$ beams where
$\sqrt{s_{e^+e^-}} = 500~{\rm GeV}$. Explicit expressions
for $f_\lambda(x)$ may be found in Ref.~\cite{gi83}. We should
emphasize that we choose this arrangement primarily for its simplicity.
We have made no attempt at optimizing the experimental arrangement since
we simply wish to investigate the degree to which the resonant dip
in the $M_{WW}$ spectrum survives in a ``typical'' setup.

Figure 3a shows the $W^+W^-$ invariant mass spectrum
for various values of the Higgs boson  mass; the spectrum corresponds to
Eq.~\ref{eq:weighted} multiplied by the square of the
branching fraction ${\rm Br}(W\rightarrow {\rm hadrons}) \simeq .7$ since
$M_{WW}$ is most reliably reconstructed by requiring
both W bosons to decay hadronically. Furthermore, we
have required the laboratory angles of both the $W^+$ and $W^-$
to obey\footnote{Due to limitations
in the computer time necessary to achieve acceptable statistics,
we have compromised by imposing lab angle cuts on the $W$ bosons
rather than their jet decay products. It should be noted that
authors sometimes impose angular cuts in the $\gamma \gamma$ center
of mass frame which do not necessarily correspond to configurations
accessible to experiments\cite{gu92,bo93,be93,ye91}.}
$| \cos\theta_{\rm lab} | < .85 .$ For comparison, the dashed line
in Fig. 3a includes only the nonresonant tree-level
background processes.

Due to imperfect mass reconstruction in an actual experiment,
the distribution of the measured invariant mass of a $W^+W^-$ pair,
$M_{WW}^{meas},$
will differ from that shown in Fig.~3a. To simulate this effect
we define ``smeared" cross sections through
convolution with a Gaussian resolution function,
\begin{equation}
{\displaystyle d\tilde\sigma \over dM_{WW}^{meas.} }
= \int~ dM_{WW}
\displaystyle{ 1 \over  \sqrt{ 2 \pi }\sigma_{res}  }
\exp
\left[
- \displaystyle{ ( M_{WW}^{meas} - M_{WW} )^2 \over 2 \sigma^2_{res}  }
\right]
{\displaystyle d \sigma \over d M_{WW}} ,
\end{equation}
where $\sigma_{res}$ is the experimental resolution.
Fig. 3b demonstrates how $\sigma_{res}=5$~GeV smooths
the details of the distribution shown Fig.~3a. The dashed line
in Fig.~3b corresponds to smearing the tree-level background
processes.

Let us investigate how the including interference effects
changes the nature of the Higgs boson signal in the $M_{WW}$ distribution.
Suppose we define the Higgs boson signal $S$ as the excess or deficit
of $W^+W^-$ pairs in a specific mass interval with respect to
tree-level background expectations,
\begin{equation}
\label{eq:signal}
S = f \times {\displaystyle \int dM_{WW}^{meas}
\left( {\displaystyle d\tilde\sigma \over dM_{WW}^{meas} }
     - {\displaystyle d\tilde\sigma^{\rm tree} \over dM_{WW}^{meas} }
\right)
     }.
\end{equation}
The factor $f$ is the $e^+e^-$ integrated luminosity
which is appropriate since we assume each $e^+$($e^-$) gives
exactly one backscattered photon. We will assume a
yearly integrated luminosity of $f = 20~{\rm fb}^{-1}$\cite{gu92,be93}
for our calculations.
Similarly, we define the background $B$ as
\begin{equation}
\label{eq:background}
B = f \times {\displaystyle \int dM_{WW}^{meas}
    {\displaystyle d\tilde\sigma^{\rm tree} \over dM_{WW}^{meas} }
     }.
\end{equation}

The nature of the interference between the
Higgs boson signal and the nonresonant background may be deduced
from Fig.~4 which plots the fractional excess of $W^+W^-$ pairs,
$S/B,$ as a function of $M_H.$
For definiteness we choose an integration interval
of $M_H \pm 2 \Gamma_H$ in
Eqs.~\ref{eq:signal},\ref{eq:background}
and assume perfect resolution
($\sigma_{res} \rightarrow 0~{\rm GeV}$ as in Fig. 3a).
The net interference effect (after integrating over
the chosen interval) is destructive beyond $m_H \gwig  200~{\rm GeV}.$
For comparison, the dashed line in Fig.~4
illustrates the inevitable fractional
excess if one (incorrectly) combines the signal and background
processes incoherently. The change of slope in both curves at
$M_H = 2 M_Z \simeq 182~{\rm  GeV}$ is due to the $H \rightarrow 2 Z$
threshold in the Higgs boson width.

We can go beyond the qualitative nature of Fig.~4 by considering the
statistical significance of a Higgs boson signal
in the $\gamma \gamma \rightarrow W^- W^+$ channel.
If we wish to employ a simple $|S|/\sqrt{B}$ measure of
statistical significance, it would clearly be
unwise to restrict our attention to
$S$ defined over a $M_{WW}$ bin centered on $M_H$ as we did for
Fig.~4 since such bins potentially encompass regions
of both constructive and destructive interference.
However, for a fixed binwidth one
can optimize the position of the bin center to maximize
$|S| /\sqrt{B};$ though such an optimization is
statistically illegal (unless one compensates for the
reduction in the number of degrees of freedom) we will nevertheless
use this technique to place an upper limit on the significance attainable.

With the binwidth chosen as $\max(2 \sigma_{res},\Gamma_H),$
the solid curve in Fig.~5 plots the maximum $|S|/\sqrt{B}$ obtainable
for $\sigma_{res} = 5~{\rm GeV}$ for the backscattered laser
facility we consider. The dashed curve in Fig.~5
gives the value of $|S| / \sqrt{B}$ obtained
in the spirit of previous analyses which a) neglect interference effects
b) assume experimental resolution effects are accounted for solely by
the binwidth (i.e., they impose no Gaussian smearing) and
c) center the bins on $M_H.$ Because of these differences the curves in
Fig.~5 are not directly comparable: the dashed curve is included only
for reference to previous studies. As in Fig.~4, there is a change in slope
of the curves in Fig.~5 at $M_H = 2 M_Z$ due to an abrupt change in
the Higgs boson decay width (neglecting threshold effects).

Our motivation for pursuing the issue of interference effects
in  $\gamma \gamma \rightarrow H \rightarrow W^- W^+$ was
the possibility that significant interference, if present,
could brighten the prospects for exploiting the $W^+W^-$
mode for detecting a Higgs boson with $M_H > 2 M_Z$ at
a backscattered laser facility. We have demonstrated
that the relevant interference terms can indeed be large and that for
$M_H \gwig 200~{\rm GeV}$ they result in a net
destructive interference which not only negates the contributions
from an incoherent signal ( i.e., from the square of the loop
amplitudes) but also produces a net resonant dip in the
corresponding $M_{WW}$ spectrum: this is the central result of this
letter. Unfortunately, the statistical significance of the resonant
dip is not encouraging (at least in the experimental arrangement considered)
and thus the  $W^+W^-$ mode for $M_H > 200~{\rm GeV}$ remains elusive even
though the underlying nature of the signal is significantly altered.

\vspace{.75cm}
We would like to thank B. Cousins and J. Hauser for
enlightening discussions. D.A.M. acknowledges the hospitality
of the S.S.C. Theory Group while part of this work was being completed.
T.N.T. likewise acknowledges the U.C.L.A. Theory Group for an enjoyable
visit. D.A.M. is supported by the
Eloisatron project. D.Z. is supported by the I.N.F.N.

\section*{ Appendix}

For completeness, we summarize
expressions for the relevant tree and loop amplitudes of Fig.~1.
We use the Feynman rules listed in Ref.~\cite{ao80} and the phase conventions
described below.

In the $W^+W^-$ center of momentum system
we define the contravariant components of the  momentum
and polarization vectors for the $W^-$ boson as
\begin{eqnarray}
\label{eq:mom} p^{\mu}
& = &(E,|\vec p|\sin\theta^* ,0,|\vec p|\cos\theta^*)\\
\label{eq:tran}
\epsilon^{\mu}_{\pm} & = &{1\over{\sqrt{2}}}(0,\cos\theta^*,
\pm i,-\sin\theta^*)\\
\epsilon^{\mu}_{0}
& = &{1\over M_W}(|\vec p|,E\sin\theta^*,0,E\cos\theta^*)
\end{eqnarray}
where $E=\sqrt{\vec p^{~2}+M_W^2}.$
The corresponding components for the $W^+$ boson
are obtained by the substitution
$\theta^* \rightarrow \pi + \theta^*.$ The components for a photon
traveling along the $+z~(-z)$ axis follow from Eqs.~\ref{eq:mom},\ref{eq:tran}
by substituting $\theta^* \rightarrow 0$ ($\theta^* \rightarrow \pi$),
where $E = |\vec{p}|$ is understood.

The sum of the amplitudes corresponding to Fig.~1 may be written as
\begin{equation}
A_{\lambda_1 \lambda_2 \lambda_3 \lambda_4}=
A_{\lambda_1 \lambda_2 \lambda_3 \lambda_4}^{\rm loop} +
A_{\lambda_1 \lambda_2 \lambda_3 \lambda_4}^{\rm tree}
\end{equation}
where $\lambda_1, \lambda_2, \lambda_3, \lambda_4$
refer to the helicities of the photon along the $+z$ axis,
the photon along the $-z$ axis, the $W^-$ boson and the $W^+$ boson
respectively.

The loop amplitudes may be expressed as
\begin{eqnarray}
\lefteqn{ A_{\lambda_1\lambda_2\lambda_3\lambda_4}^{\rm loop} =
-{{\alpha g^2}\over {8\pi}}~~{{
(\epsilon_{\lambda_1}\cdot\epsilon_{\lambda_2})
(\epsilon^*_{\lambda_3}\cdot\epsilon^*_{\lambda_4}) s}
\over{s-M^2_H+i\Gamma_H M_H}}  \times } \nonumber \\
& &
\left[  3r_W (2-r_W) G(r_W) + 3r_W + 2
- 2 \sum_f Q^2_f r_f \left\{ (1-r_f) G(r_f) + 1 \right\} \right]
\end{eqnarray}
where $r_i = 4 M_i^2 / s$ and the sum is over all massive fermions
species (with charge $Q_f$ in units of $|e|$), including color as a
degree of freedom. The function
$G(r)$ is given by
\begin{equation}
G(r) = \left\{
\begin{array}{ll}
 \displaystyle{ \pi^2 \over 4 }
-\displaystyle{   1   \over 4 }
 \ln^2 \left( \displaystyle{ 1+\sqrt{1-r} \over 1-\sqrt{1-r} } \right) +
 \displaystyle{ i \pi \over 2}
 \ln \left( \displaystyle{ 1+\sqrt{1-r} \over 1-\sqrt{1-r} } \right)
& \qquad \qquad r<1 \nonumber \\
 &  \\
 \left( \arctan \displaystyle{ 1 \over \sqrt{r-1} }  \right)^2
 &\qquad \qquad  r > 1 .\nonumber
\end{array}
\right.
\end{equation}

In our phase conventions the tree amplitudes which
interfere with the loop amplitudes may be written as
\begin{equation}
A^{\rm tree}_{\lambda_1\lambda_2\lambda_3\lambda_4}
=
{\displaystyle 8 \pi \alpha \over 1 - \cos^2\theta^* + r_W \cos^2\theta^* }
\times
\tilde A^{\rm tree}_{\lambda_1\lambda_2\lambda_3\lambda_4} ,
\end{equation}
where
\begin{equation}
\label{eq:treeamps}
\tilde{A}^{\rm tree}_{++\pm\pm} = 2 \pm 2 \sqrt{ 1 - r_W } - r_W,
\qquad \qquad \tilde{A}^{\rm tree}_{++00} = r_W,
\end{equation}
with $\tilde A^{\rm tree}_{\lambda_1\lambda_2\lambda_3\lambda_4}
=\tilde A^{\rm tree}_{-\lambda_1-\lambda_2-\lambda_3-\lambda_4}.$
All other tree amplitudes do not interfere with the loop amplitudes
and hence their overall phases are irrelevant for our purposes.
Explicit expressions for them may be found in Ref.~\cite{be92}.

\newpage
\section*{Figure Captions}
\begin{itemize}

\item
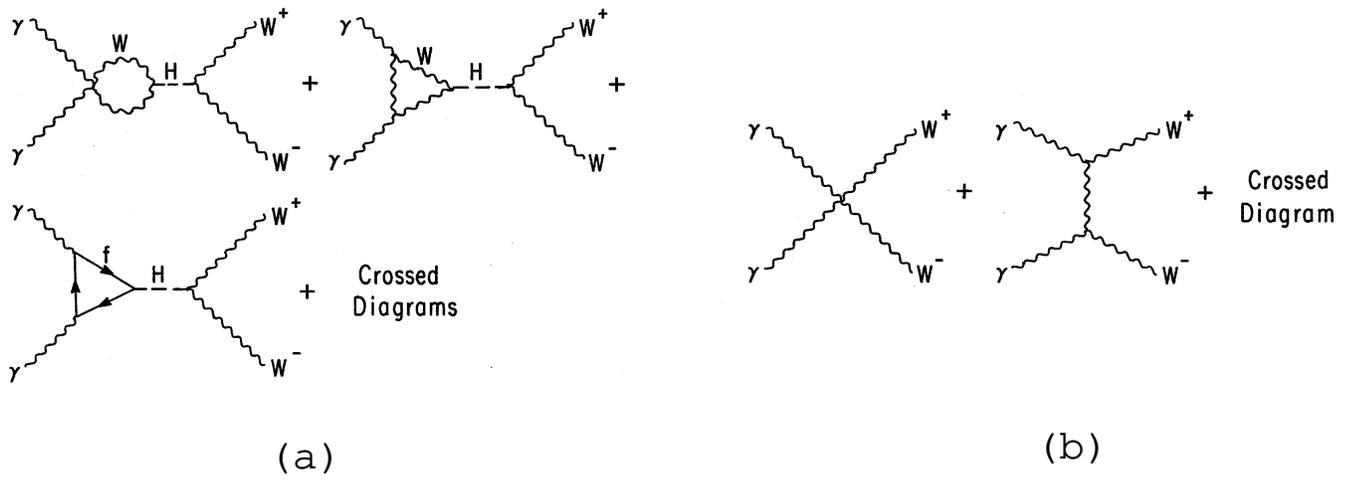
[{\bf Fig. 1}] a) One loop diagrams for
$\gamma \gamma \rightarrow H \rightarrow W^+ W^-.$  b) Nonresonant
tree-level diagrams for $\gamma \gamma \rightarrow W^+ W^-.$

\item
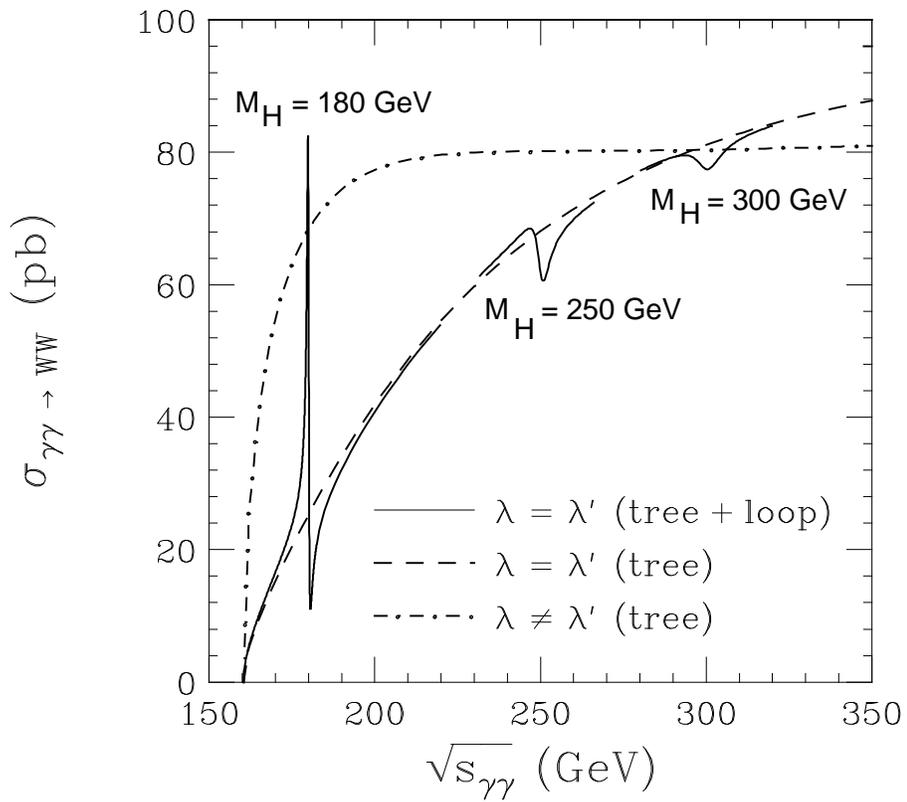
[{\bf Fig. 2}]
Photon-photon cross sections for $W$ pair production as a function
of $\gamma \gamma$ center of mass energy $\sqrt{s_{\gamma\gamma}}$
and incoming photon helicities $\lambda,\lambda'.$  For three values
of the Higgs boson mass we show cross sections which
have been integrated over all $W^+W^-$ scattering angles and summed over
$W$ boson polarizations.
Solid lines include interference
effects between tree and loop amplitudes of Fig.~1.
Dashed and dot-dashed lines include only
tree level contributions.

\item
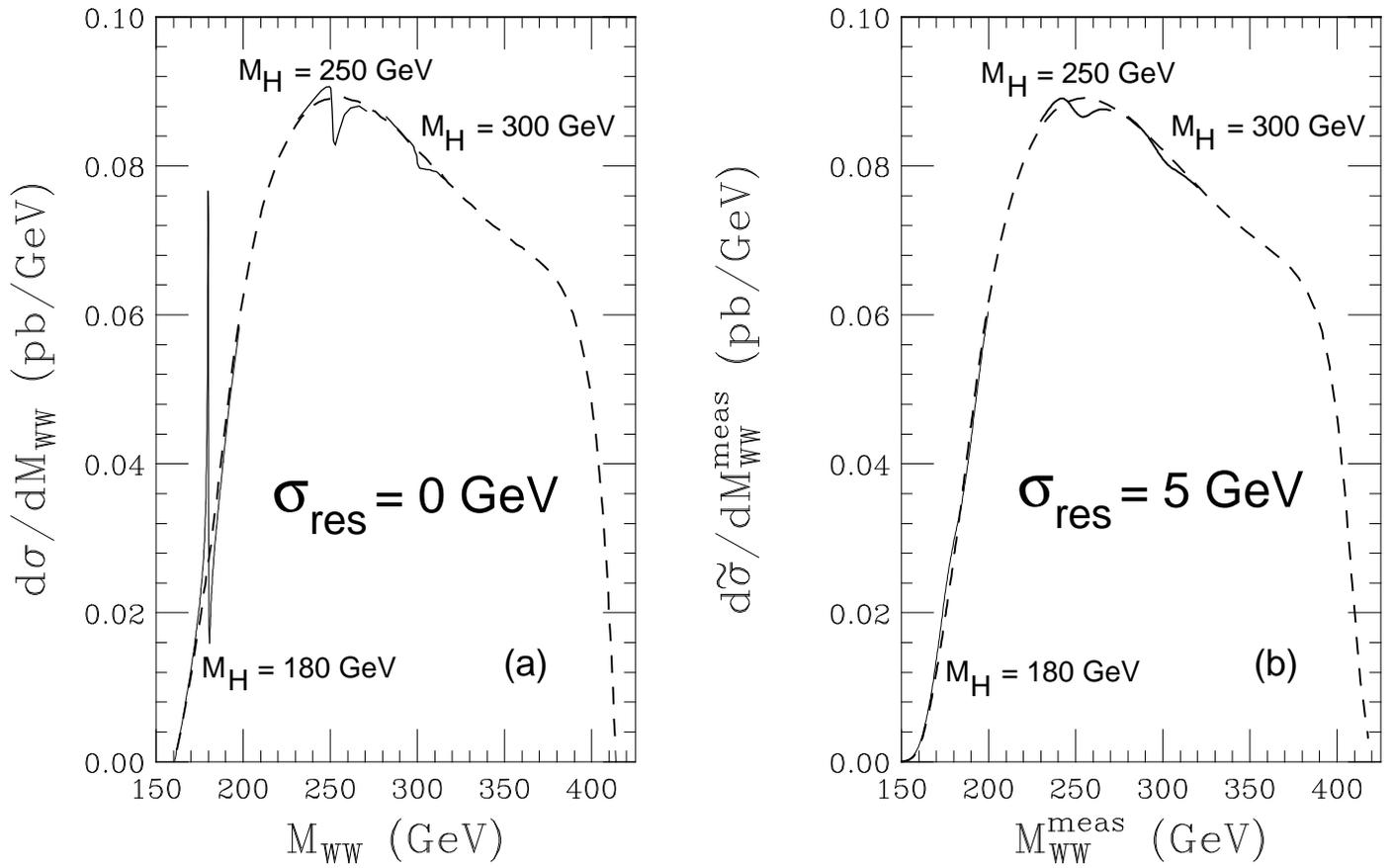
[{\bf Fig. 3}]
(a) $W^+W^-$ invariant mass spectrum at a
$\sqrt{s_{e^+e^-}} = 500~{\rm GeV}$ backscattered laser facility
with completely polarized lasers ($\lambda^{\rm laser}_1 =
\lambda^{\rm laser}_2$) and completely unpolarized $e^+(e^-)$ beams.
Solid lines include interference between tree and loop amplitudes.
Dashed line includes only nonresonant tree amplitudes. Perfect
experimental resolution is assumed. (b) Same as
in (a) except with $\sigma_{\rm res}=5~{\rm GeV}$ experimental
resolution effects implemented.

\item
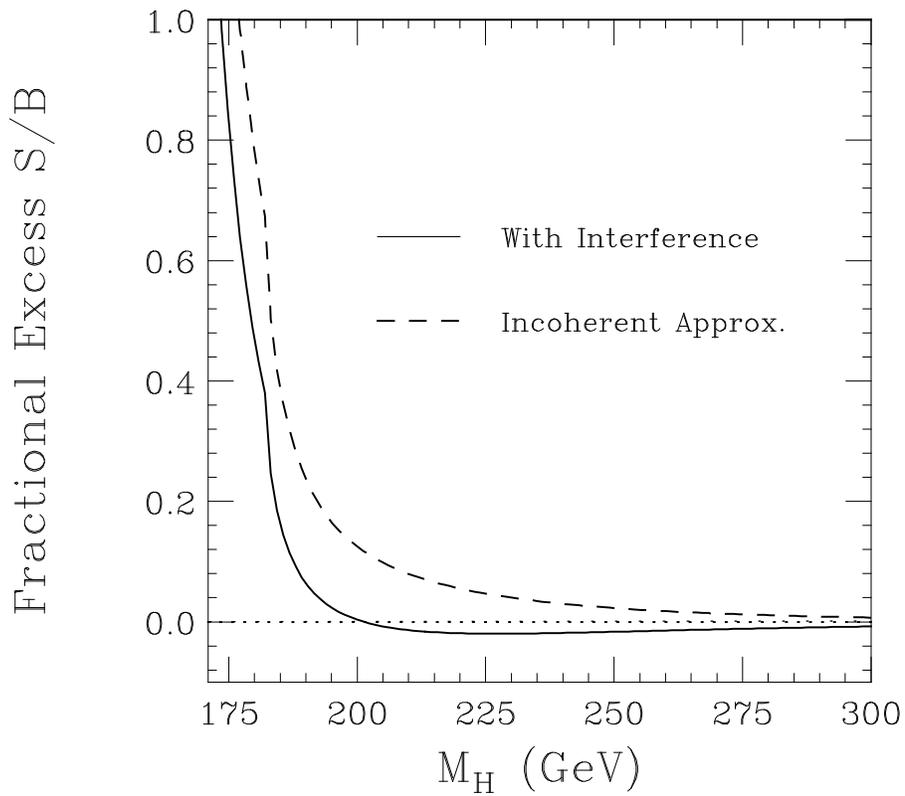
[{\bf Fig. 4}]
Fractional excess of $W^+W^-$ pairs in the $M_{WW}$ interval
$M_H \pm 2 \Gamma_H$ assuming perfect experimental resolution
at a $\sqrt{s_{e^+e^-}} = 500~{\rm GeV}$ backscattered laser facility
with completely polarized lasers ($\lambda^{\rm laser}_1 =
\lambda^{\rm laser}_2$) and completely unpolarized $e^+(e^-)$ beams.
Solid line includes interference between tree and loop amplitudes
whereas dashed line neglects interference  effects. The horizontal dotted
line at $S/B = 0$ suggests that net interference effect is destructive
for $m_H \gwig 200~{\rm GeV}.$

\item
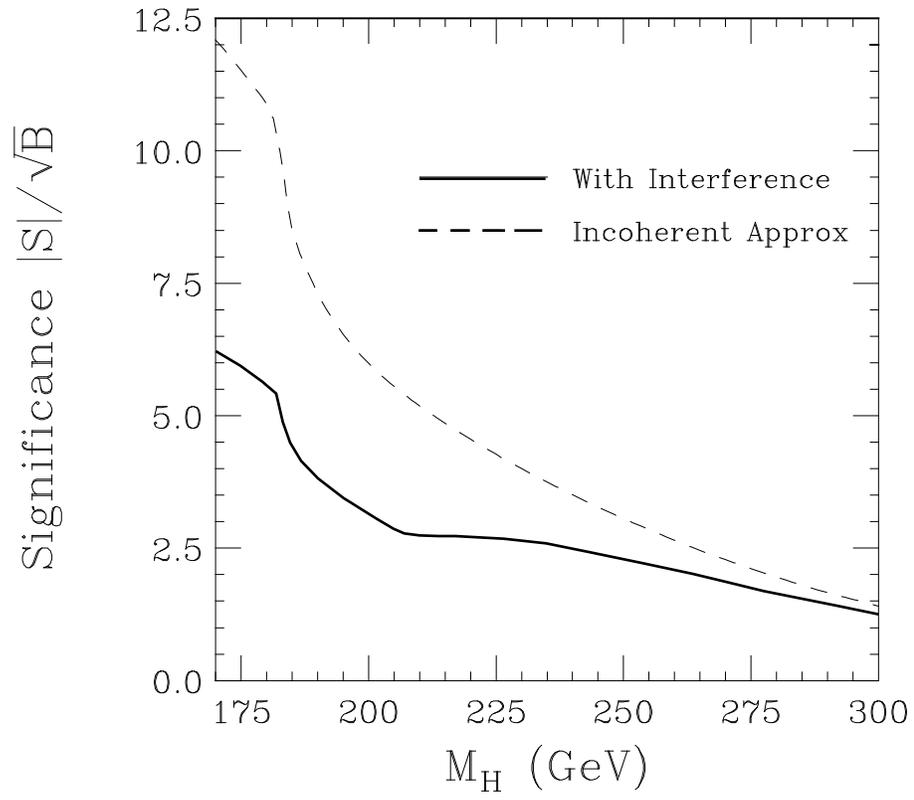
[{\bf Fig. 5}]
Statistical significance of Higgs boson signal in the
$W^+W^-$ invariant mass spectrum at a
$\sqrt{s_{e^+e^-}} = 500~{\rm GeV}$ backscattered laser facility
with completely polarized lasers ($\lambda^{\rm laser}_1 =
\lambda^{\rm laser}_2$) and completely unpolarized $e^+(e^-)$ beams.
An integrated $e^+e^-$ luminosity of $20~{\rm fb}^{-1}$ is assumed.
The binwidth for both curves is fixed at
$2\sigma_{\rm res} = 10~{\rm GeV}.$  The solid curve includes
interference effects, $\sigma_{\rm res} = 5~{\rm GeV}$
resolution effects, and corresponds to an optimized bin center.
The dashed curve neglects interference effects and assumes resolution
effects are accounted for solely by the binwidth with bins
centered on $M_H.$
\end{itemize}

\begin{thebibliography}{99}

\bibitem{gu92}
J.F. Gunion and H.E. Haber, UCD preprint, UCD-92-22, August 1992.
\bibitem{bow93}
D. Boswer-Chao and K. Cheung, Phys Rev. {\bf D48}, (1993) 89.
\bibitem{bo93}
D.L. Borden, D.A. Bauer, D.O. Caldwell, UCSB preprint UCSB-HEP-93-01,
March 1993.
\bibitem{gi83}
I.F. Ginzburg, G.L. Kotkin, V.G. Serbo and V.I. Telnov, Nucl. Inst. Meth.
{\bf 205}, (1983), 47;
I.F. Ginzburg, G.L. Kotkin, S.L. Panfil,
V.G. Serbo and V.I. Telnov, Nucl. Inst. Meth. {\bf 219}, (1984), 5.
\bibitem{tel90}
V.I. Telnov, Nucl. Instr. Meth. {\bf A294}, (1990) 72.
\bibitem{ji93}
G.V. Jikia, Phys. Lett. {\bf B298} (1993) 224.
\bibitem{be93}
M. Berger, UW Madison preprint MAD/PH/771, July 1993.
\bibitem{di93}
D. Dicus and C. Kao, U. of Texas preprint CCP-93-24, August 1993.
\bibitem{ve93}
H. Veltman, private communication to T.N.T.
\bibitem{va79}
A.I. Vainshtein, M.B. Voloshin, V.I. Zakharov and M.A. Shifman,
Sov. J. Nucl. Phys. {\bf 30}, (1979), 711 [Yad. Fiz. {\bf 30}, (1979), 1368].
\bibitem{ye91}
E. Yehudai, Phys. Rev. {\bf D44}, (1991), 3434.
\bibitem{be92}
G. Belanger and F. Boudjema, Phys. Lett. {\bf B288}, (1992), 210.
\bibitem{ao80}
K. Aoki {\it et al.}, Supp. Prog. Theor. Phys., {\bf 73} (1982) 1.
\end{thebibliography}
\end{document}